\documentclass[12pt]{article}
\usepackage{amsmath,amssymb,bm,epsfig}
\usepackage{cancel}

\renewcommand{\thefootnote}{\fnsymbol{footnote}}

\begin{document}

\title{
\begin{flushright}
\begin{minipage}{0.2\linewidth}
\normalsize
WU-HEP-15-11 \\*[50pt]
\end{minipage}
\end{flushright}
{\Large \bf 
Particle physics and cosmology with high-scale SUSY breaking in five-dimensional supergravity models 
\\*[20pt]}}

\author{Hajime~Otsuka\footnote{
E-mail address: hajime.13.gologo@akane.waseda.jp
}\\*[20pt]
{\it \normalsize 
Department of Physics, Waseda University, 
Tokyo 169-8555, Japan} \\*[50pt]}

\date{
\centerline{\small \bf Abstract}
\begin{minipage}{0.9\linewidth}
\medskip 
\medskip 
\small
We discuss a high-scale SUSY breaking scenario with the wino dark matter in the five-dimensional supergravity model on $S^1/Z_2$. 
The extra $U(1)$ symmetries broken by the orbifold projection control the flavor structure of soft SUSY-breaking parameters as well 
as the Yukawa couplings, and a scalar component of the one of moduli multiplets, which arise from extra-dimensional components of the $U(1)$ vector multiplets, induces the slow-roll inflation. 
Because of the supersymmetric moduli stabilization as well as the moduli inflation, it is found that the correct dark matter relic abundance is non-thermally generated by the gravitino decaying into the wino. 
\end{minipage}
}

\begin{titlepage}
\maketitle
\thispagestyle{empty}
\clearpage
\tableofcontents
\thispagestyle{empty}
\end{titlepage}

\renewcommand{\thefootnote}{\arabic{footnote}}
\setcounter{footnote}{0}

\section{Introduction}
\label{sec:introduction}
The standard model (SM) is the successful theory which describes the interactions of elementary particles, in agreement with all the experimental results below the electroweak (EW) scale. 
However, there are several unsolved theoretical and observational problems such as the mass hierarchy of elementary particles, 
non-existence of the viable dark matter candidates and even the gravitational interactions in the SM. 

When we consider the extension of the SM, there are two important ideas indicated from the string theory which is expected as the consistent theory of quantum gravity. 
One of them is the supersymmetry (SUSY). 
From the phenomenological point of view, the supersymmetric models are also attractive scenario beyond the SM, because the minimal supersymmetric standard model (MSSM) gives 
rise to the gauge coupling unification at a high energy scale around $2.1\times 10^{16}$ GeV, stable SUSY particles as dark matter candidates, and so on. 
In addition, the local supersymmetric models, i.e., supergravity models describe the gravitational interactions in the framework of four-dimensional supergravity ($4$D SUGRA). 
The other idea is the extra dimension. In the extra-dimensional models, the hierarchical Yukawa couplings arise from the wavefunction profiles of quarks, 
leptons and Higgs, propagating in the bulk of the extra-dimensional space. (See for more details, e.g., Ref.~\cite{ArkaniHamed:1999dc} in the five-dimensional cases.) 

Based on these ideas, we consider the five-dimensional supergravity model ($5$D SUGRA) which is the minimal extension of the SM with local supersymmetry and a single extra dimension. 
Among them, $5$D SUGRA compactified on $S^1/Z_2$ is the simplest but a workable theory, where the four dimensional chirality arises from the orbifolding of fifth dimension. 
Although there are several systematic studies focusing on the particle phenomenology~\cite{Abe:2008an} or cosmology~\cite{Paccetti:2005zm,Abe:2014vca} based on 
the off-shell formulation of $5$D SUGRA~\cite{Zucker:1999ej,Kugo:2000af}, it would be quite important to treat them on the same footing and perform a complementary study. 
For such a purpose, in this paper we propose the realistic $5$D SUGRA model, consistent with the recent LHC experiments~\cite{Chatrchyan:2014lfa} as well as the Planck results~\cite{Ade:2015lrj}. 

Before discussing the details of the $5$D SUGRA, we review the recent implications from the LHC experiments. 
The LHC Run I results indicate no evidence for SUSY up to a scale of ${\cal O}(1)$ TeV ~\cite{Chatrchyan:2014lfa}, and the observed Higgs boson mass 
$m_h\simeq 126$ GeV~\cite{Aad:2012tfa} is considerably larger than that expected in the framework of MSSM with a low-scale SUSY breaking, 
although a few possibilities are pointed out to raise the Higgs boson mass within the low-scale SUSY scenarios~\cite{Abe:2007kf,Feng:1999zg}. 
As a class of models compatible with the observed Higgs boson mass in the MSSM, the high-scale SUSY breaking scenarios are often considered, 
such as Split SUSY~\cite{Wells:2003tf}, Spread SUSY~\cite{Hall:2011jd} and Pure gravity mediation~\cite{Ibe:2011aa} and so on. 
In these scenarios, the gaugino masses are relatively lighter than the soft SUSY-breaking masses. For example, in the case of pure gravity mediation, 
the gaugino masses are determined by the anomaly mediation~\cite{Randall:1998uk,Giudice:1998xp}, whereas the other soft SUSY-breaking terms are determined by the gravity mediation. 
Then, the lightest supersymmetric particle (LSP) is the wino.~\cite{Randall:1998uk,Giudice:1998xp}\footnote{The higgsino dark matter is also considered in the pure gravity mediation scenario~\cite{Evans:2014pxa}.}. 
However, the relic abundance of wino depends on the thermal history of the universe after the inflation. 
Especially, the non-thermal productions of the wino LSP is highly model-dependent. Indeed, the wino LSP is produced by the decays of gravitino~\cite{Gherghetta:1999sw}, moduli~\cite{Moroi:1999zb} and Q-ball~\cite{Fujii:2001xp} after the inflation. 
Therefore, we discuss such high-scale SUSY breaking models with the wino dark matter based on the concrete model, i.e., $5$D SUGRA on $S^1/Z_2$ in which the successful inflation mechanism can be realized~\cite{Abe:2014vca}. 

The paper is organized as follows. In Sec.~\ref{subsec:2_1}, we briefly review the matter contents and their interactions in our model derived from the $5$D SUGRA. 
Then, the moduli stabilization, inflation dynamics and SUSY breaking mechanism are shown in Sec.~\ref{subsec:2_2}. 
Sec.~\ref{sec:3} is devoted to the thermal history of the universe in our model, and then the relevant decay processes into the wino dark matter are discussed. 
We find that the relic abundance of the wino is determined by the non-thermal productions from the gravitino decay, whereas the other decay process via the inflaton decay 
is not relevant to the present dark matter abundance. 
Finally, Sec.~\ref{sec:con} is devoted to the conclusion.

\section{Setup}
\label{sec:2}
\subsection{Four-dimensional effective action on $S^1/Z_2$}
\label{subsec:2_1}
In this section, we now to proceed to details of the $5$D SUGRA on $S^1/Z_2$ with flat $5$D spacetime whose coordinates 
are denoted as $x^M=(x^\mu, y)$ with $\mu=0,1,2,3$ and $y$ is fifth coordinate within the range $0\leq y \leq L$. 
The matter contents in this background are classified as two types of fields, such as $Z_2$-even and $Z_2$-odd fields, 
where only $Z_2$-even fields are appeared in the low energy effective theory after the dimensional reduction of the fifth dimension. 

First of all, we show the moduli chiral multiplets involved in $5$D $Z_2$-odd $U(1)_{I'}$ vector multiplets ${\bm V}^{I'}$. 
Since the $5$D SUSY is broken into the $4$D ${\cal N}=1$ SUSY by the $Z_2$-orbifolding, these vector multiplets are 
decomposed into the $4$D $Z_2$-odd vector multiplets $V^{I'}$ and $Z_2$-even chiral multiplets $\Sigma^{I'}$, i.e., ${\bm V}^{I'}=\{ V^{I'}, \Sigma^{I'}\}$ 
with $I'=1,2,\cdots n_{V^{I'}}$. 
Only the zero-modes of $\Sigma^{I'}$ can be appeared in the low-energy effective theory below the compactification scale, whereas $V^{I'}$ do not have zero-modes. 
We denote the zero-modes of $\Sigma^{I'}$ as $T^{I'}$ whose imaginary parts of scalar components correspond to the fifth components of $U(1)_{I'}$ gauge fields $A_M^{I'}$, 
and then $T^{I'}$ are called as the moduli chiral multiplets. 
For $5$D $Z_2$-even vector multiplets ${\bm V}^I$ and hypermultiplets ${\bm H}_\alpha$, they are also decomposed into the $4$D $Z_2$-even vector multiplets 
$V^I$ and chiral multiplets $\Sigma^I$, $\Phi_\alpha$ and $\Phi_\alpha^C$, i.e., ${\bm V}^I=\{V^I, \Sigma^I\}$ and ${\bm H}_\alpha=\{\Phi_\alpha, \Phi_\alpha^C\}$ 
with $I=1,2,\cdots, n_{V^I}$ and $\alpha=1,2,\cdots n_H+1$. Here, we assume the single compensator multiplet, for simplicity. 
As for these zero-modes of $V^I$ and $\Phi_\alpha$, we further assume the following MSSM matter contents plus three right-handed (s)neutrino multiplets $N_i$, 
stabilizer multiplets $H_a$ ($a=1,2,\cdots, n_{H_a})$ and SUSY breaking multiplet $X$;
\begin{eqnarray}
 (V_1,V_2,V_3) &:& \mbox{gauge vector multiplets}, \nonumber\\
 ({\cal Q}_i,{\cal U}_i,{\cal D}_i) &:& \mbox{quark chiral multiplets}, \nonumber\\
 ({\cal L}_i,{\cal E}_i, N_i) &:& \mbox{lepton chiral multiplets}, \nonumber\\
 ({\cal H}_u,{\cal H}_d) &:& \mbox{Higgs chiral multiplets},\nonumber\\
 (H_a) &:& \mbox{stabilizer chiral multiplets},\nonumber\\
 (X) &:& \mbox{SUSY breaking chiral multiplet},
 \label{eq:multi}
\end{eqnarray} 
where $V_1$, $V_2$, $V_3$ denote the gauge multiplets for $U(1)_Y$, $SU(2)_L$, $SU(3)_C$, and $i$ denotes the flavor index. 
As shown later, $H_a$ and $X$ are responsible for the moduli stabilization and the spontaneous SUSY breaking, respectively. 
It is remarkable that these chiral multiplets originating from the hypermultiplets ${\bm H}_\alpha$ have extra $U(1)_{I'}$ charges $c_{I'}^{(\alpha)}$ under the $U(1)_{I'}$ gauge symmetries. 

From the $5$D conformal supergravity action~\cite{Zucker:1999ej,Kugo:2000af}, 
we can obtain the effective K\"ahler potential and gauge kinetic function after the off-shell dimensional reduction~\cite{Abe:2005ac,Correia:2006pj,Abe:2006eg} 
in terms of the $4$D ${\cal N}=1$ superspace~\cite{Abe:2004ar,Paccetti:2004ri}. (See Refs.~\cite{Sakamura:2011df,Sakamura:2012bj} for more general description of $5$D SUGRA including the $Z_2$-odd fields.) 
Then, $4$D effective Lagrangian becomes
\begin{align}
 {\cal L}_{\rm eff} =& -\frac{1}{4}\left[\int d^2\theta ~ 
 \sum_r f_r(T)\text{tr}({\cal W}^r{\cal W}^r )+\text{h.c.} \right] +\int d^4\theta ~~ |\phi|^2 \Omega_{\rm eff}(|Q|^2,\text{Re} T)
 \nonumber \\ 
 &+\left[\int d^2\theta ~~ \phi^3 W(Q,T)+\text{h.c.} \right], 
 \label{eq:Leff}
\end{align}
where $\phi$ is the compensator multiplet, ${\cal W}^r$ are the field strength supermultiplet for 
$4$D gauge multiplets $V^r$ with $r=1,2,3$, $Q_{\alpha}$ denote all the $4$D chiral multiplets given by Eq.~(\ref{eq:multi}). 

Such $5$D action is characterized by the cubic polynomial of vector multiplets, 
\begin{align}
{\cal N}(M) =\sum_{{I,J,K}=1}^{n_V} C_{I,J,K}M^IM^JM^K,
\end{align}
where $C_{I,J,K}$ is the real constant. If the 5D supergravity models 
are embedded in the more higher-dimensional theories, such as 
heterotic M-theory, these coefficients $C_{I,J,K}$ correspond to 
the intersection numbers of the internal Calabi-Yau manifolds~\cite{Lukas:1998yy}. 
By identifying the real coefficients $C_{I',J,K}$ with $\xi_{I'}^r$, the gauge kinetic functions $f_r(T)$ in Eq.~(\ref{eq:Leff}) are represented as
\begin{equation}
 f_r(T) =\sum_{I'=1}^{n_{V^{I'}}}\xi^r_{I'}T^{I'}.
 \label{eq:gaugekin}
\end{equation}
In addition, the effective K\"ahler potential in Eq.~(\ref{eq:Leff}) is 
\begin{align}
 \Omega_{\rm eff}(|Q|^2,\text{Re} T) 
 &= -3{\cal N}^{1/3}(\text{Re} T)\left[
 1 -\frac{2}{3}\sum_a Y(c^{(\alpha)} \cdot T) |Q_\alpha|^2 +\sum_{\alpha, 
\beta}\tilde{\Omega}^{(4)}_{\alpha ,\beta}(\text{Re} T) 
|Q_\alpha|^2 |Q_\beta|^2 
 +{\cal O}\Bigl( |Q|^6\Bigl) \right], \nonumber\\
 \label{eq:effKahler}
\end{align}
where $c_{I'}^{(\alpha)}$ denote the $U(1)_{I'}$ charges of $Q_\alpha$ whose kinetic term is given by
\begin{align}
Y(z)\equiv \frac{1-e^{-2\text{Re} z}}{2\text{Re} z},
\end{align}
where the exponential factor can be extracted from the exponential form of the localized wavefunctions propagating in the bulk, controlled by the bulk mass, i.e., $U(1)_{I'}$ charges in our framework. 
Moreover, the four-point couplings $\tilde{\Omega}^{(4)}_{\alpha, \beta}$ are defined as
\begin{eqnarray}
 \tilde{\Omega}^{(4)}_{\alpha ,\beta} &\equiv& \frac{(c^{(\alpha)}
\cdot{\cal P} a^{-1}\cdot c^{(\beta)})
 \{ Y((c^{(\alpha)} +c^{(\beta)})\cdot T)-Y(c^{(\alpha)}\cdot T)
Y(c^{(\beta)}\cdot T) \} }
 {3(c^{(\alpha)} \cdot \text{Re} T)(c^{(\beta)} \cdot \text{Re} T)}
 -\frac{Y((c^{(\alpha)} +c^{(\beta)})\cdot T)}{9}, \nonumber\\ 
\label{eq:Omg4}
\end{eqnarray}
where ${\cal P}^{I}_{~~J}\equiv \delta^{I}_{~~J}-{\cal X}^I{\cal N}_J/3{\cal N}$ project the moduli multiplets out the radion multiplet which corresponds to the single modulus $T^{I'=1}$ in the case $n_{V^{I'}}=1$. 
In contrast to the K\"ahler potential, the superpotential is only allowed at the orbifold fixed points $y=0,L$ due to its holomorphicity. 

\subsection{Brief review of moduli inflation}
\label{subsec:2_2}
Along the line of Refs.~\cite{Abe:2014vca} and \cite{Otsuka:2015oma}, 
we briefly review the supersymmetric moduli inflation which cannot be realized in the single modulus case. 
In order to obtain the Yukawa hierarchies between generations as shown in Sec.~\ref{sec:3}, we consider the case of three moduli multiplets, 
and at the same time, three stabilizer multiplets are introduced to generate the moduli potential, i.e., $n_{V^{I'}}=n_{H_a}=3$. 
The norm function for the moduli multiplets is chosen as 
\begin{equation}
{\cal N} (\text{Re} T) =(\text{Re} T^1)(\text{Re} T^2)(\text{Re} T^3),
\end{equation}
for simplicity. 
In addition to the bulk configurations in the $5$D SUGRA on $S^1/Z_2$ given by Eq.~(\ref{eq:effKahler}), we set the following superpotential at the orbifold fixed points $y=0,L$,
\begin{align}
{\cal W}=J_0^{(a)} {\cal H}_a\delta( y)-J_L^{(a)} {\cal H}_a \delta( y-L),
\end{align}
where $J_{0,L}^{(a)}$ are real constants and ${\cal H}_a$ are the stabilizer chiral multiplets. 
Here, we assume that such linear terms at the fixed points are dominant terms protected by certain symmetries and dynamics. 
By integrating the $5$D action over the fifth dimension~\cite{Abe:2005ac,Correia:2006pj,Abe:2006eg}, we find the following superpotential $W$, 
\begin{align}
W= (J_0^{(a)} -J_L^{(a)} e^{-c_{I'}^{(a)} T^{I'}}) H_a,
\label{eq:esp}
\end{align}
and the K\"ahler potential for the moduli and stabilizer fields, 
\begin{equation}
K= -{\rm ln} {\cal N} ({\rm Re} T)+ \sum_{a=1}^3 Z_{H_a}(\text{Re}T^{I'=a}) 
|H_a|^2, 
\label{eq:Kmo}
\end{equation}
where 
\begin{align}
Z_{H_a}(\text{Re}T^{I'=a})=\frac{1-e^{-2c^{(a)}_{I'}{\rm Re}T^{I'}}}{c^{(a)}_{I'} {\rm Re}T^{I'}}.
\label{eq:esp}
\end{align}
It is then assumed that the stabilizer fields $H_a$ have only $U(1)_{I'=a}$ charge, for simplicity. 

Next, let us discuss the moduli stabilization from the scalar potential in $4$D ${\cal N}=1$ SUGRA 
by employing the above K\"ahler and superpotential,\footnote{Here and hereafter, we employ the reduced Planck mass unit, i.e., $M_{\rm Pl}=1$.} 
\begin{align}
V&= e^K (K^{m\bar{n}} D_m WD_{\bar{n}} \bar{W} -3|W|^2),
\label{eq:sp}
\end{align}
where $D_m W=W_m +K_m W$, $W_m=\partial_m W$, $K_m=\partial_m K$, 
$K_{m\bar{n}}=\partial_m\partial_{\bar{n}} K$, 
$K^{m\bar{n}}=(K^{-1})_{m\bar{n}}$, $m,n=\{I', a\}$, and at the moment, we omit the contributions from the other fields expect for the moduli and stabilizer chiral multiplets, $T^{I'}, H_a$. 
The extremal conditions of them $\langle V_{T^{I'}}\rangle=\langle V_{H_a}\rangle =0$ with 
$V_m =\partial_m V$ are satisfied under the supersymmetric conditions, i.e., $\langle D_{T^{I'}}W \rangle=\langle D_{H_a}W\rangle=0$ which lead to their vacuum expectation values (VEVs),
\begin{align}
c_{I'}^{(a)} \langle T^{I'}\rangle = \ln \left( \frac{J_L^{(a)}}{J_0^{(a)}}\right),
\,\,\langle H_a\rangle =0,
\label{eq:movac}
\end{align}
and the vanishing cosmological constant $\langle V\rangle=0$.\footnote{Similar moduli stabilizations are discussed in the single modulus case~\cite{Maru:2003mq} 
and in the case of multi compensators~\cite{Abe:2007zv}.}
At the extrema given by Eq.~(\ref{eq:movac}), moduli $T^{I'}$ 
and stabilizer fields $H_a$ have same supersymmetric masses, 
\begin{align}
m_{I'a}^2\simeq \frac{e^{\langle K\rangle }
\langle W_{I'a}\rangle^2}{K_{I'\bar{I}'} K_{a\bar{a}}},
\label{eq:THmass}
\end{align}
where $W_{I'a}=-c_{I'}^{(a)}J_L^{(a)}e^{-c_{I'}^{(a)}T^{I'}}$, $W_{ij}=\partial_i\partial_j W$, and the real and imaginary parts of them also have the same masses at the vacuum. 
From the mass formula~(\ref{eq:THmass}), we can consider that the one pair of modulus and stabilizer fields, e.g. ($T^{I'=1}$, $H_{a=1}$) is lighter enough than the other 
pairs ($T^{I'\neq 1}$, $H_{a\neq 1}$) by choosing $|c_{I'\neq 1}^{(a\neq 1)}| < |c_{I'= 1}^{(a=1)}|$ and $|J_{0,L}^{(1)}|< |J_{0,L}^{(a\neq 1)}|$ in Eq.~(\ref{eq:esp}). 
In such cases, the pairs ($T^{I'\neq 1}$, $H_{a\neq 1}$) are fixed at their supersymmetric minimum given by Eq.~(\ref{eq:movac}), and then the only remaining 
pair ($T^{I'=1}$, $H_{a=1}$) appears in the low energy effective action given by the following K\"ahler and superpotential, 
\begin{align}
&K= -\ln {\rm Re} T^1+ Z_{H_1}(\text{Re}T^{1})|H_1|^2 +\cdots, \nonumber\\
&W= (J_0^{(1)} -J_L^{(1)} e^{-c_1^{(1)} T^{1}}) H_1,
\label{eq:KWinf}
\end{align}
where the ellipsis stands for the VEVs of the other moduli fields $T^{I'\neq 1}$. 
By employing the $4$D scalar potential~(\ref{eq:sp}) replaced with Eq.~(\ref{eq:KWinf}), the effective potential for the modulus $T^1$ becomes
\begin{align}
V_{\rm inf} &= e^K K^{H_1\bar{H}_1} |W_{H_1}|^2 \simeq 
\frac{|J_{0}^{(1)} -J_{L}^{(1)}e^{-c_1^{(1)} T^1}|^2}{\langle{\rm Re}\,T^2
\rangle \langle{\rm Re}\,T^3
\rangle (1-e^{-c_1^{(1)}T^1})},
\label{eq:infpo}
\end{align}
on the $H_1=0$ hypersurface. 
When the inflaton is identified with the real part of modulus ${\rm Re}\,T^1$, our obtained effective scalar potential on the ${\rm Im}\,T^1=0$ hypersurface is similar 
to the one in Starobinsky model~\cite{Starobinsky:1980te}. 
During the inflation, $H^1$ and ${\rm Im}\,T^1$ are fixed at the origin due to the Hubble-induced mass and real parameters $J_{0,L}^{(a)}$, respectively. 
By solving the equation of motion for ${\rm Re}\,T^1$ with the following parameter settings,
\begin{align}
&J_{0}^{(2)}=J_{0}^{(3)}=\frac{1}{9},\,\,
J_{L}^{(2)}=J_{L}^{(3)}=1,\,\,
J_{0}^{(1)}=\frac{1}{4000},\,\,J_{L}^{(1)}=\frac{3}{4000},\,\,
&c_2^{(2)}=c_3^{(3)}=\frac{1}{50},\,\,\,
c_1^{(1)}=\frac{1}{10}, 
\label{eq:para}
\end{align}
the cosmological observables such as the power spectrum of the scalar curvature perturbation $P_\xi (k)$, its spectral index $n_s$, 
and the tensor-to-scalar ratio $r$ with an enough amount of e-foldings $N\simeq 58$ are obtained,
\begin{align}
P_\xi (k_\ast)\simeq 2.2\times 10^{-9},\,\,\,
n_s \simeq 0.96,\,\,\,
r \simeq 10^{-5},
\end{align}
at the pivot scale $k_\ast=0.05\,[{\rm Mpc}^{-1}]$. These results are consistent with the recent Planck data, reported in Ref.~\cite{Ade:2015lrj}, 
\begin{align}
P_\xi (k_\ast) \simeq 2.20\pm 0.10 \times 10^{-9},\,\,\,
n_s = 0.9655\pm 0.0062,\,\,\,
r < 0.11. 
\end{align}

The supersymmetric masses of moduli and stabilizer fields are also given by 
\begin{align}
m_{T^2}\simeq m_{T^3}\simeq m_{H_2}\simeq m_{H_3}\simeq 
4.8\times 10^{15}\,{\rm GeV},\,\,\,
m_{T^1}\simeq m_{H_1}\simeq 4\times 10^{12}\,{\rm GeV},
\label{eq:momass}
\end{align}
with the numerical values of the parameters~(\ref{eq:para}), which shows that the fields except for the pair ($T^{I'=1}$, $H_{a=1}$) are decoupled from the inflaton dynamics as mentioned before. 
It turns out that the typical Kaluza-Klein mass is also larger than the inflaton mass, 
\begin{equation}
M_{C} =\frac{\pi}{L} \simeq \frac{\pi}{\langle {\cal N}^{1/2}\rangle} \simeq 2.1\times 10^{16} \text{GeV},
\end{equation}
with the moduli VEVs,
\begin{equation}
(\text{Re} \langle T^1\rangle, \text{Re} \langle T^2\rangle, \text{Re} \langle T^3\rangle ) \simeq (11,\;110,\;110),
\label{eq:movev}
 \end{equation}
by utilizing the parameters given by Eq.~(\ref{eq:para}).
So far, the F-terms of moduli and stabilizer fields vanish at the vacuum, because there are no source of SUSY breaking in the current setup.

For the SUSY breaking sector,  as an example, we take an O'Raifeartaigh model~\cite{O'Raifeartaigh:1975pr} 
determined by the following K\"ahler and superpotential of the SUSY breaking field $X$, 
\begin{align}
K=Z_X(\text{Re}T^2, \text{Re}T^3) |X|^2 
-\cfrac{1}{\Lambda^2} |X|^4,\,\,\,
W=w +\nu X,
\label{eq:KWX}
\end{align}
where $w$, $\nu$ are the real parameters and the loop corrections in the K\"ahler potential are characterized by the scale of heavy mode or 
dynamical scale $\Lambda$~\cite{Kallosh:2006dv}. 
It is then assumed that the SUSY breaking field $X$ have no charge for $U(1)_1$ gauge symmetry, which forbids the inflaton decay into the SUSY breaking field at the classical level. 

As discussed in Refs.~\cite{Abe:2014vca} and \cite{Otsuka:2015oma}, when the SUSY breaking scale is enough lighter than the inflation mass, 
the moduli and stabilizer fields remain stabilized at the values close to their minimum~(\ref{eq:movac}) during the inflation. In fact, the deviations from them and VEV of $X$,
\begin{align}
&\delta H_a \simeq \frac{m_{3/2}}{m_{H_a}},\,\,\, \delta T^{I'}\simeq \left(\frac{m_{3/2}}{m_{T^{I'}}}\right)^2, 
X \simeq -\cfrac{\Lambda^2 (Z_X)^2 w}{4\nu} \ll 1,
\label{eq:dev}
\end{align}
are small enough, where $m_{3/2}=e^{\langle K\rangle/2}\langle W\rangle$. 
Furthermore, the F-terms of these fields are estimated at their minimum involving the deviations~(\ref{eq:dev}) from the supersymmetric minimum~(\ref{eq:movac}), 
\begin{align}
&\sqrt{K_{T^{I'}{\bar T}^{I'}}}F^{T^{I'}} \simeq O\left(\frac{m_{3/2}^3}{m_{T^{I'}}^2}\right), \;\;  \sqrt{K_{H_a\bar{H}_a}}F^{H_{a}} \simeq O\left(\frac{m_{3/2}^3}{m_{H_a}^2}\right),\;\; \sqrt{K_{X\bar{X}}}F^X \simeq \frac{-\nu}{\langle {\cal N}\rangle^{1/2}\langle Z_X\rangle^{1/2}},
\label{eq:Fterm} 
\end{align}
from which the F-terms of moduli and stabilizer fields are suppressed by the gravitino mass. 
Thus, one can discuss the low-and high-scale SUSY breaking scenarios, as long as such scales are lower than the inflation scale. 

\section{Wino dark matter in the high-scale SUSY breaking}
\label{sec:3}
In this section, we discuss the particle phenomenology of $5$D SUGRA based on the moduli inflation in Sec.~\ref{subsec:2_2}. 
Since the moduli do not induce the SUSY breaking at the vacuum, the sizable gaugino masses are not generated at the tree-level as seen in Eq.~(\ref{eq:gaugekin}). 
Throughout this paper, we do not consider the gauge kinetic function at the boundary fixed point $y=0$, such as $f^{(0)}=\xi_xX$ with $\xi_x$ being the real parameter, 
from which the F-term of $X$ induces the sizable gaugino masses, although the R-symmetry is explicitly broken.\footnote{Such situation is discussed in Refs.~\cite{Otsuka:2015oma}.} 
Thus, the dominant contributions to the gaugino masses come from the anomaly mediation~\cite{Randall:1998uk,Giudice:1998xp}, 
\begin{align}
M_r&=\frac{b_rg_r^2}{16\pi^2}m_{3/2},
\label{eq:gauginoano}
\end{align}
where the VEV of the conformal compensator, $\langle F^\phi \rangle/\langle \phi\rangle$ is replaced by the gravitino mass $m_{3/2}$ 
due to the almost vanishing F-terms of the moduli and the stabilizer fields~(\ref{eq:Fterm}) and tiny VEV of $X$ (\ref{eq:dev}), 
$\gamma_\alpha^\beta=\frac{1}{16\pi^2}\left(\frac{1}{2}y^{\alpha \beta \gamma}y^{\ast}_{\alpha \beta \gamma} 
-2g_r^2C_r(S_\alpha)\delta^\alpha_\beta\right)$ are the anomalous dimension with $S_\alpha$ being the fields in the MSSM, 
$C_r(S_\alpha)$ are the quadratic Casimir invariants, 
$b_r=(33/5,1,-3)$ are the beta function coefficients in the MSSM. 
As pointed out in Refs.~\cite{Giudice:1998xp,Gherghetta:1999sw}, there are the threshold corrections to the gaugino masses at the one-loop level. 
Especially, when the $\mu$-term is of order the gravitino mass $m_{3/2}$, the heavy higgsino threshold corrections give rise to
\begin{align}
M_r&=\frac{b_rc_r g_r^2}{16\pi^2}
\mu\,{\rm sin}2\beta \frac{m_A^2}{|\mu|^2 -m_A^2}
\ln \frac{|\mu|^2}{m_A^2},
\end{align}
in the limit of $m_W \ll \mu, m_A$, where $c_r=(1/11,1,0)$, ${\rm tan}\,\beta=v_d/v_u$ with $v_{u,d}$ being the VEVs of up-and down-type Higgs fields, 
$m_W$ and $m_A$ are the masses of W-boson and CP-odd heavy Higgs boson, respectively. The origin of the $\mu$-and $B\mu$-terms is shown later. 

By contrast, the soft SUSY-breaking masses are generated via the four point couplings between $X$ and the matter multiplets $S_\alpha$ in the MSSM~(\ref{eq:Omg4}). 
Although they depend on the $U(1)_{I'}$ charge of $S_\alpha$, typical sparticle masses are of order the gravitino mass which are larger than the gaugino masses by one-loop factor~(\ref{eq:gauginoano}). 
It turns out that these spectra are similar to those of pure gravity mediation~\cite{Ibe:2011aa}. 
In such cases, the LSP is the wino by the nature of renormalization group (RG) effects, and the thermal abundance of wino is determined by the wino mass, i.e., gravitino mass, 
whereas the non-thermal abundance of wino is highly model-dependent. 
Therefore, we investigate these abundances by focusing on the wino productions from the gravitino decay based on our moduli stabilization as well as the inflation. 
These gravitinos are produced by the thermal bath and scalar field ($T^{I'}, H_a, X$) decays after the inflation in Sec.~\ref{subsec:2_2}. 
In the following, we discuss the details of the sparticle spectra and the relic abundance of wino step by step.

\subsection{Yukawa hierarchies and Sparticle spectra} 
\label{subsec:3_1}
So far, we discuss the inflaton and SUSY breaking sectors. 
To complete our discussion, we set the superpotential for the fields in the MSSM and (s)neutrinos. 
For the matter chiral multiplets involved in the hypermultiplets $\Phi_\alpha$, we put the following Yukawa couplings at the boundary fixed point $y=0$, 
\begin{equation}
W_{\rm MSSM} = \lambda_{ij}^u{\cal Q}_i{\cal H}_u{\cal U}_j
 +\lambda_{ij}^d{\cal Q}_i{\cal H}_d{\cal D}_j +\lambda_{ij}^e{\cal L}_i{\cal H}_d{\cal E}_j +\lambda_{ij}^n{\cal L}_i{\cal H}_u N_j, 
\label{eq:wmssm}
\end{equation}
where $\lambda_x^{ij}$ ($x=u,d,e,n$) are the holomorphic Yukawa couplings. 
We assign the R-charge to the chiral multiplets in Eq.~(\ref{eq:multi}) as $R_X=R_{H_a}=2$, $R_{{\cal Q}_i}=R_{{\cal U}_i}=R_{{\cal D}_i}=R_{{\cal L}_i}=R_{{\cal E}_i}=R_{N_i}=1$, $R_{{\cal H}_u}=R_{{\cal H}_d}=0$, 
where $R_{\phi}$ is the R-charge of $\phi$. 
When the matter chiral multiplets are canonically normalized, the physical Yukawa couplings are expressed as 
\begin{align}
y_u^{ij}&=\frac{\lambda_u^{ij}}{\sqrt{\hat{Y}_{{\cal Q}_i}\hat{Y}_{{\cal H}_u}\hat{Y}_{\bar{{\cal U}}_j}}},\,\,
y_d^{ij}=\frac{\lambda_d^{ij}}{\sqrt{\hat{Y}_{{\cal Q}_i}\hat{Y}_{{\cal H}_d}\hat{Y}_{\bar{{\cal D}}_j}}},\,\,
y_e^{ij}=\frac{\lambda_e^{ij}}{\sqrt{\hat{Y}_{{\cal L}_i}\hat{Y}_{{\cal H}_d}\hat{Y}_{\bar{{\cal E}}_j}}},\,\,
y_n^{ij}=\frac{\lambda_n^{ij}}{\sqrt{\hat{Y}_{{\cal L}_i}\hat{Y}_{{\cal H}_u}\hat{Y}_{\bar{N}_j}}},
\end{align}
where 
\begin{align}
\hat{Y}_{\alpha}=\partial_{\alpha}\partial_{\bar{\alpha}}\Omega 
={\cal N}^{1/3}({\rm Re}T)\left( 2 Y(c^{(\alpha)}\cdot T)-3\Omega_{\alpha, X}|X|^2 +{\cal O}(|X|^4) \right).
\end{align}
With the $U(1)_{I'}$ charge assignments of the matter 
multiplets as summarized in Table~\ref{tab:charge} and the ${\cal O}(1)$ value of the holomorphic Yukawa couplings $\lambda_x^{ij}$ ($x=u,d,e,n$), 
the observed quark and lepton masses and their mixing angles are realized as shown in Ref.~\cite{Otsuka:2015oma}.\footnote{By changing the R-charges of fields and introducing the mass terms of the 
Majorana neutrinos, the tiny masses of the neutrinos are also explained. These terms are put on the superpotential at the other boundary fixed point $y=L$, such as $W=\kappa_{ij}e^{-2c_{I'}^{N_i}\hat{T}^{I'}}N_iN_j$, 
where $\kappa_{ij}$ are ${\cal O}(1)$ parameters and $c_{I'}^{(N_i)}$ ($i=1,2,3$) are the $U(1)_{I'=i}$ charge of the Majorana neutrino chiral multiplets $N_i$ for the linear combination 
of the vector multiplets $V^{I'}$. Such terms induce the non-thermal leptogenesis~\cite{Asaka:1999yd} via the non-thermal inflaton decay into the Majorana neutrino, i.e., $T^3\rightarrow N_3N_3$ as 
also discussed in Ref.~\cite{Czerny:2014xja}. } 
Here, the theoretical values are evaluated in terms of the one-loop RG equations of the MSSM from the EW scale to the compactification scale $M_C$. \begin{table}[h]
\begin{center}
\begin{tabular}{|l|l|l|} \hline
\rule[-2mm]{0mm}{7mm}
$c^{{\cal Q}_i}_{I'=3}=(0.1,~0.1,~1.1)
$ & $c^{{\cal L}_i}_{I'=3}=(0.1,~0.1,~1.6)$ & $c^{{\cal H}_u}_{I'=3}=0
$\\ 
$c^{{\cal Q}_i}_{I'=2}=(-0.1,-0.1, 0.8)$ 
& $c^{{\cal L}_i}_{I'=2}=(-0.1,-0.1,~0)$ & $c^{{\cal H}_u}_{I'=2}=0.1$ \\ 
$c^{{\cal Q}_i}_{I'=1}=(0.1,~0.4,~1)$ 
& $c^{{\cal L}_i}_{I'=1}=(0.1,~0.5,~0)$ & $c^{{\cal H}_u}_{I'=1}=-0.9$ \\ \hline
\rule[-2mm]{0mm}{7mm}
$c^{{\cal U}_i}_{I'=3}=(0.1,~0.1,~0.6)
$ & $c^{{\cal E}_i}_{I'=3}=(0.1,~0.2,~0.2)$ & $c^{{\cal H}_d}_{I'=3}=0
$\\ 
$c^{{\cal U}_i}_{I'=2}=(-0.1,-0.1, 0.3)$ 
& $c^{{\cal E}_i}_{I'=2}=(-0.1,-0.1,0)$ & $c^{{\cal H}_d}_{I'=2}=0$ \\ 
$c^{{\cal U}_i}_{I'=1}=(-0.2,~0.2,~1)$ 
& $c^{{\cal E}_i}_{I'=1}=(-0.2,~0,~-0.5)$ & $c^{{\cal H}_d}_{I'=1}=-0.1$ \\ \hline
\rule[-2mm]{0mm}{7mm}
$c^{{\cal D}_i}_{I'=3}=(0.1,~0.1,~0.2)
$ & $c^{N_i}_{I'=3}=(0.1,~0.1,~0.1)
$ & \\ 
$c^{{\cal D}_i}_{I'=2}=(-0.1,-0.1,0)$ 
& $c^{N_i}_{I'=2}=(-0.3,-0.3,-0.3)$ &   \\ 
$c^{{\cal D}_i}_{I'=1}=(0.3,~0.2,~-0.5)$ & $c^{N_i}_{I'=1}=(-0.7,~-0.7,~-0.7)$ 
&   \\ \hline
\end{tabular}
\caption{$U(1)_{I'}$ charges of the quarks, leptons and Higgs.}
\label{tab:charge}
\end{center}
\end{table}

In addition, we consider the following superpotential as the counterpart of $\mu$-term, 
\begin{equation}
W_{\mu-{\rm term}} =\sum_{a=1}^3\kappa_a H_a {\cal H}_u{\cal H}_d, 
\end{equation}
where $\kappa^a$ ($a=1,2,3$) are the real parameters, and then the VEVs of $H_a$ generate the effective $\mu$-term, 
\begin{align}
\mu =\sum_{a=1}^3\frac{\kappa_a \langle H_a\rangle}
{\sqrt{\hat{Y}_{H_a}\hat{Y}_{{\cal H}_u}\hat{Y}_{{\cal H}_d}}},
\end{align}
where the relevant fields are canonically normalized. 
By assuming that $\mu$-term is only generated by the VEV of $H_3$, i.e., $\kappa_1=\kappa_2=0$, the scale of $\mu$-term is of order the gravitino mass which is required for the 
successful EW symmetry breaking in high-scale SUSY breaking scenario,
\begin{align}
\mu \simeq 5.7\times 10^{-3} \kappa_3 \frac{m_{3/2}}{m_{H_1}}M_{\rm Pl} 
\simeq {\cal O}(m_{3/2}),
\end{align} 
where the $U(1)_{I'}$ charges of relevant fields given by Eq.~(\ref{eq:para}) and Tab.~\ref{tab:charge}, the VEVs of $T^{I'}$ (\ref{eq:movev}) and the mass of stabilizer field (\ref{eq:momass}) are employed. 
Note that the mild large volume of the fifth dimension $y$ reduces the scale of $\mu$-term by the factor $5.7\times 10^{-3}$, and furthermore the existence of this three-point couplings do not 
depend on the dynamics of inflaton and stabilizer fields thanks to the tiny VEVs of Higgs fields. 
In order to realize the successful EW symmetry breaking, we add the Giudice-Masiero terms~\cite{Giudice:1988yz} to the K\"ahler potential at the boundary fixed point $y=0$ 
as the origin of $\mu$-and $B\mu$-term,
\begin{align}
K^{(0)}=c {\cal H}_u{\cal H}_d +{\rm h.c.},
\end{align} 
where $c$ is a constant and the mass scales of $\mu$-and $B\mu$-terms are close to the gravitino mass. 
These holomorphic terms cannot be derived from the bulk K\"ahler potential~(\ref{eq:effKahler}) which respects the ${\cal N}=2$ SUSY.   

The $U(1)_{I'}$ charge assignments of the matter multiplets determine not only the observed masses and mixing angles of the elementary particles, 
but also the flavor structure of the soft SUSY-breaking terms through the four point couplings between the SUSY breaking field $X$ and the matter multiplets in the effective K\"ahler potential~(\ref{eq:Omg4}). 
They are characterized as
\begin{align}
{\cal L}_{\rm soft} =&-\sum_{\alpha}m_{S_\alpha}^2 |S_\alpha|^2 
-\left(\frac{1}{2}\sum_{r=1}^3 M_r \lambda^r\lambda^r 
+\frac{1}{6}\sum_{\alpha, \beta, \gamma}a_{\alpha \beta \gamma}S_\alpha S_\beta S_\gamma +B\mu h_uh_d 
+{\rm h.c.}\right),
\end{align}
where $S_\alpha=h_u,h_d,\tilde{q}_i,\tilde{u}_i,\tilde{d}_i,\tilde{l}_i,\tilde{e}_i,
\tilde{\nu}_i$ and the scalar components of ${\cal H}_u,{\cal H}_d,{\cal Q}_ i,{\cal U}_i,
{\cal D}_i,L_i,{\cal E}_i,N_i$ and $\lambda^r$ ($r=1,2,3$) are 
the gauginos. These soft terms except for the gaugino masses are given by the well known formulas~\cite{Choi:2005ge,Kaplunovsky:1993rd}, 
\begin{align}
m_{S_\alpha}^2= -\langle F^I\bar{F}^{\bar{J}} \partial_I \partial_{\bar{J}} \ln (\hat{Y}_{S_\alpha})\rangle,
\,\,\,
B &\simeq -m_{3/2},
\,\,\,
a_{\alpha \beta \gamma}=y_{\alpha \beta \gamma}\langle F^I\partial_J
\ln (\hat{Y}_{S_\alpha }\hat{Y}_{S_\beta}\hat{Y}_{S_\gamma})\rangle,
\label{eq:softterms} 
\end{align}
where the indices $I$ and $J$ run over all the chiral multiplets. 
Since the moduli and the stabilizer fields do not induce the SUSY breaking effects, the gaugino masses and the 
A-terms are almost vanishing. 
Therefore, the leading contribution to them are the one-loop anomaly mediated effects which are expressed as Eq.~(\ref{eq:gauginoano}) for gaugino masses and 
\begin{align}
a_{\alpha \beta \gamma}&=-\left(\gamma^\zeta_\alpha y_{\zeta \beta \gamma}+\gamma^\zeta_\beta y_{\alpha \zeta \gamma}
+\gamma^\zeta_\gamma y_{\alpha \beta \zeta}\right) m_{3/2},
\end{align}
for A-terms. 
On the other hand, the anomaly mediation for the soft SUSY-breaking masses can be negligible relative to the gravity mediation for them. 
It should be noted that there are the experimental constraints to the wino mass. 
Since the unstable gravitino have to decay before the Big-Bang Nucleosynthesis (BBN), the wino mass is constrained to $M_2> 200-250\,{\rm GeV}$. 
The LHC experiments also give the lower bound for the wino mass $M_2> 270\,{\rm GeV}$ by searching for the disappearing tracks~\cite{Aad:2013yna}.

\subsection{The relic abundance of wino dark matter}
Next, we discuss the relic abundance of the wino LSP from the gravitino decays. 
The gravitino is produced by the thermal bath and non-thermal decays of scalar fields such as the inflaton Re\,$T^1$, 
real parts of lightest stabilizer field Re\,$H_1$ and SUSY breaking field Re\,$X$ after the inflation. 
The other moduli and stabilizer fields are decoupled from the inflaton dynamics and fixed at their minimum. 
Especially, the imaginary parts of fields, Im\,$T^1$, Im\,$H_1$ and Im\,$X$ do not oscillate around their minimum after the inflation, because we set the real parameters in the 
superpotential~(\ref{eq:esp}) and (\ref{eq:KWX}). 
Thus, we focus on the non-thermal productions of the wino LSP given through the decays of Re\,$T^1$, Re\,$H_1$, Re\,$X$ after the inflation. 
As summarized in Ref.~\cite{Otsuka:2015oma}, the gravitino yield $Y_{3/2}=n_{3/2}/s$ where $n_{3/2}$ is the number density of gravitino and $s$ is the entropy density of universe is estimated.
First, from the inflaton decay into the gravitino(s), the gravitino yield is given by
\begin{align}
Y_{3/2}^{\sigma^1} &\simeq \text{Br}(\sigma^1 \rightarrow  \Psi_\text{3/2}\Psi_\text{3/2}) \frac{3T_R}{4m_{T^1}} \simeq \frac{1}{288\pi \langle K_{T^1\bar{T}^1} \rangle \Gamma_{\rm all}^{\sigma^1}} \frac{3m_{3/2}^2T_R}{4M_{\rm Pl}^2} \simeq 2\times 10^{-19} \left( \frac{m_{3/2}}{10^5\,{\rm GeV}}\right)^2 \left( \frac{T_R}{10^9\,{\rm GeV}}\right),
\end{align}
where $\sigma^1={\rm Re}\,T^1$, $\Gamma_{\rm all}^{\sigma^1}\simeq 3.95\,{\rm GeV}$ is the total decay width of inflaton which is obtained from the inflaton decay into the gauge boson pairs. 
The reheating temperature $T_R$ is estimated by equaling the expansion rate of the universe, $H_R=H(T_R)$ to $\Gamma_{\rm all}^{\sigma^1}$,
\begin{align}
T_{R} = \left( \cfrac{\pi^2 g_\ast (T_R)}{90}\right)^{-1/4} \sqrt{\Gamma_{\rm all}^{\sigma^1} M_{\rm Pl}} \simeq  1.38\times 10^9~{\rm GeV},
\label{eq:totreh}
\end{align}
where $g_\ast (T_R) =915/4$ is the effective degrees of freedom in the MSSM at the reheating $T_R$. 

Next, the gravitino yield is produced by the $h_1={\rm Re}\,H_1$ decay,
\begin{align}
Y_{3/2}^{h_1} &=\frac{2 \rho_{h_1}}{m_{H_1}s}\simeq 
 \frac{1}{4}\frac{m_{3/2}^2 T_R}{m_{H_1}^3} =2.5\times 10^{-18}\left( \frac{m_{3/2}}{10^5\,{\rm GeV}}\right)^2\left( \frac{10^{12}\,{\rm GeV}}{m_{H_1}}\right)^3\left( \frac{T_R}{10^9\,{\rm GeV}}\right),
\end{align}
where $\rho_{S_\alpha}$ is the energy density of $S_\alpha$. 
Finally, the SUSY breaking field $x={\rm Re}\,X$ also produces the gravitino,
\begin{align}
Y_{3/2}^{x} &\simeq 
\frac{3}{2}\frac{T_R}{m_X}
\left(\frac{m_{3/2}}{m_X}\right)^{16/3}
\left( \frac{\Gamma_{\rm all}^{\sigma^1}}
{\Gamma_{\rm all}^{x}}\right)^{2/3} \simeq 1.2\times 10^{-18}\left( \frac{m_{3/2}}{10^5\,{\rm GeV}}\right)^{20/3}\left( \frac{10^{9}\,{\rm GeV}}{m_{X}}\right)^{29/3}\left( \frac{T_R}{10^9\,{\rm GeV}}\right),
\end{align}
where $\Gamma_{\rm all}^{x}$ is the total decay width of $x$, 
\begin{align}
\Gamma_{\rm all}^x\simeq 
\Gamma (x \rightarrow \Psi_{3/2}\Psi_{3/2})
\simeq \frac{1}{96\pi} \cfrac{m_{X}^5}{m_{3/2}^2M_{\rm Pl}^2}.
\label{eq:decx}
\end{align}

As a result, the gravitino productions from $\sigma^1$ and $h_1$ decay after the inflation are suppressed due to the almost vanishing F-terms of $T^1$ and $H_1$. 
However, the gravitinos are sufficiently produced by the decays of $x$ for the particular values of $m_X$ and $m_{3/2}$.\footnote{The details of them are also discussed in Ref.~\cite{Evans:2013nka}.}
Note that we focus on the situation where the coherent oscillation of inflaton field dominates the energy density of the universe after the inflation, and then inflaton release the entropy and reheat the 
universe at its decay. 

Moreover, the gravitino is also produced by the thermal bath in the era of radiation domination and it decay into the wino hereafter. 
The thermal abundance of gravitino is given~\cite{Bolz:2000fu,Kawasaki:2004qu,Pradler:2006qh,Pradler:2006hh} in terms of dimensionless Hubble parameter $h$ and 
the numerical parameters $y_r$, $k_r$ defined in~\cite{Pradler:2006hh},
\begin{equation}
Y_{3/2}^{\rm th} = \sum_{r=1}^3 y_r g_r(T_R)^2 \left(1+\cfrac{M_r(T_R)^2 }{3m_{3/2}^2} \right) {\rm ln}\left(\cfrac{k_r}{g_r(T_R)} \right)\times
\left(\cfrac{T_R}{10^{10} {\rm GeV}} \right),
\label{eq:thermal}
\end{equation}
where $M_r(T_R)$ and $g_r(T_R)$ are the gaugino masses and gauge couplings at the reheating $T_R$. 
The non-thermal abundance of the wino LSP is determined by solving the Boltzmann equation~\cite{Moroi:1999zb,Fujii:2001xp},
\begin{equation}
Y_{\tilde{\chi}_1^0}^{\rm nth} \simeq {\rm min}\Bigl[ Y_{3/2}^{\rm th} +Y_{3/2}^{\sigma^1}+Y_{3/2}^{h_1}+Y_{3/2}^{x}, \sqrt{\frac{45}{8\pi^2 g_\ast (T_{3/2})}}\frac{1}{M_{\rm Pl}T_{3/2}\langle \sigma_{\rm ann}v\rangle} \Bigl], 
\end{equation}
where $\tilde{\chi}_1^0$ denotes the wino-like neutralino and the entropy release from the gravitino decay is neglected. 
$g_\ast (T_{3/2})\simeq 10.75$ is the effective degrees of freedom in the MSSM at decay temperature of the gravitino,
\begin{equation}
T_{3/2}=\left( \frac{10}{\pi^2 g_\ast (T_{3/2})} M_{\rm Pl}^2 \Gamma_{3/2}^2\right)^{1/4},
\end{equation}
and $\Gamma_{3/2}$ is the gravitino decay width into the gauginos,
\begin{equation}
\Gamma_{3/2}=\frac{12}{32\pi } \frac{m_{3/2}^3}{M_{\rm Pl}^2}.
\end{equation}
It is then supposed that the sparticles except for the gauginos are heavier than the gravitino as can be seen in Tab.~\ref{tab:spectra}. 
$\langle \sigma_{\rm ann}v\rangle$ is the thermally averaged annihilation cross section of the wino-like neutralino which is roughly estimated as
\begin{align}
\langle \sigma_{\rm ann}v\rangle =\frac{3g_2^4}{16\pi M_2^2}, 
\end{align} 
in the limit of $m_W\ll M_2$, $g_2$ is the gauge coupling of $SU(2)_L$ at the EW scale. (See for more details, e.g., Ref.~\cite{ArkaniHamed:2006mb}.)
Because of the large cross section of wino, the annihilation of the wino after the non-thermal production is negligible. 
Therefore, the non-thermal production of wino is approximately given by $Y_{\tilde{\chi}_1^0}^{\rm nth} \simeq Y_{3/2}^{\rm th} +Y_{3/2}^{\sigma^1}+Y_{3/2}^{h_1}+Y_{3/2}^{x}$. 

On the other hand, the thermal abundance of the wino LSP is roughly estimated as,
\begin{align}
Y_{\tilde{\chi}_1^0}^{\rm th} \simeq \left( \sqrt{\frac{8\pi^2 g_\ast (T_{\tilde{\chi}_1^0})}{45}} \langle \sigma_{\rm ann} v\rangle M_{\rm Pl} T_{\tilde{\chi}_1^0}\right)^{-1}
\end{align} 
where $g_\ast (T_{\tilde{\chi}_1^0})\simeq 80$ is the effective degree of freedom at the freeze out temperature of wino-like neutralino $T_{\tilde{\chi}_1^0}$. 
As pointed out in Ref.~\cite{Hisano:2006nn}, the thermal abundance of the wino LSP depends on the non-perturbative effects, and then the observed dark matter density is 
realized for $m_{\tilde{\chi}_1^0} \simeq 2.7\,{\rm TeV}$ in the case of pure thermal wino. 

\begin{figure}[t]
\centering \leavevmode
\includegraphics[width=85mm]{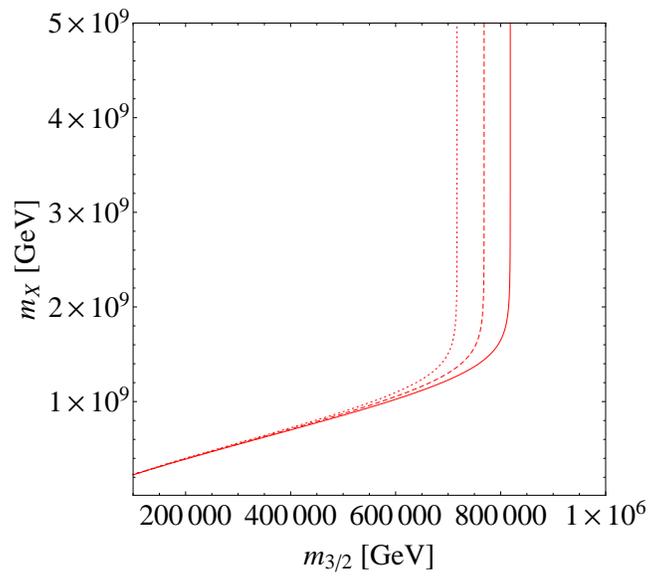}
\caption{Contours of the relic abundance of the wino LSP, $\Omega_{\tilde{\chi}_1^0} h^2=0.1$ (dotted), $\Omega_{\tilde{\chi}_1^0} h^2=0.11$ (dashed) and 
$\Omega_{\tilde{\chi}_1^0} h^2=0.12$ (solid) as a functions of the gravitino mass $m_{3/2}$ and the mass of SUSY breaking field $m_X$.}
\label{fig:wino}
\end{figure}

As a result, the relic density of the wino LSP is approximately given by the sum of thermal and non-thermal abundances, 
that is, $Y_{\tilde{\chi}_1^0}\simeq Y_{\tilde{\chi}_1^0}^{\rm th}+Y_{\tilde{\chi}_1^0}^{\rm nth}$. 
Since $Y_{\tilde{\chi}_1^0}$ becomes constant of time at low enough temperature, the total relic density of the wino is given by
\begin{equation}
\Omega_{\tilde{\chi}_1^0} h^2 = \Omega_{\tilde{\chi}_1^0}^{\rm th} h^2 +\Omega_{\tilde{\chi}_1^0}^{\rm nth} h^2,
\end{equation}
where $h\simeq 0.673$ is the present Hubble constant in units of $100$km/sec/Mpc~\cite{Ade:2015lrj}, and 
\begin{equation}
\Omega_{\tilde{\chi}_1^0}^{\rm th,nth}=m_{\tilde{\chi}_1^0} Y_{\tilde{\chi}_1^0}^{\rm th,nth}\frac{s_{\rm now}}{\rho_{\rm cr}},
\end{equation}
with $\rho_{\rm cr}$ and $s_{\rm now}$ being the critical density and the present entropy density of the universe, respectively. 
Their ratio is given by $\rho_{\rm cr}/s_{\rm now} \simeq 3.6\,h^2 \times 10^{-9}$\,GeV~\cite{Ade:2015lrj}. 
In Fig.~\ref{fig:wino}, the total relic density of the wino LSP is drawn as functions of $m_{3/2}$ and $m_X$, and then the dark matter 
abundance is constrained within the range~$0.1175 \le \Omega_{\tilde{\chi}_1^0} h^2 \le 0.1219$ reported by the Planck collaboration~\cite{Ade:2015lrj}. 
Although we do not consider the non-perturbative effects for the annihilation cross sections of wino~\cite{Hisano:2006nn}, 
we expect that the discussion is not altered within the mass range, $m_{\tilde{\chi}_1^0}\leq 1\,{\rm TeV}$. 
Even if the Sommerfeld effect significantly reduce the thermal abundance of wino-like neutralino in the mass region above ${\cal O}(1)\,{\rm TeV}$, 
the non-thermal abundance of it compensate the relic abundance of it. 

When we set the gravitino mass $m_{3/2}=1.4 \times 10^5\,{\rm GeV}$ and $m_X=2.9 \times 10^8\,{\rm GeV}$, where the total relic 
abundance of the wino is consistent with the Planck data~\cite{Ade:2015lrj}, the sparticle spectra as well as the Higgs boson mass are estimated 
as those in Table~\ref{tab:spectra} by choosing ${\rm tan}\,\beta=4$ and the $U(1)_{I'}$ charge in Tab.~\ref{tab:charge}. 
Now, the parameters $w,\nu, \Lambda$ in the superpotential~(\ref{eq:KWX}) are properly chosen to realize the above situation. 
We employ the one-loop RG equations of MSSM from the SUSY breaking scale $m_{3/2}$ to the compactification scale $M_C$, 
whereas the Higgs boson mass is evaluated by employing the result~\cite{Haber:1996fp}. 
From the Tab.~\ref{tab:spectra}, the sparticles except for the wino and chargino are of order the gravitino mass which lead to no flavor and CP-problems peculiar to the supergravity models. 

In this paper, we focus on the relic abundance of the wino LSP in the concrete $5$D SUGRA model on $S^1/Z_2$ where the successful inflation can be realized. 
The wino dark matter can be checked by the $100$ ${\rm fb}^{-1}$ data at LHC $14$ TeV running~\cite{Low:2014cba} and cosmological observations such as the 
observation of the cosmic rays from the Fermi Gamma-Ray Space Telescope~\cite{Cohen:2013ama}, AMS-$02$ experiment~\cite{Ibe:2011aa,Hall:2012zp,Hryczuk:2014hpa}, and so on. 
If the large mass of wino is prohibited by such collider experiments as well as the cosmological observations, the non-thermal process 
is important to realize the correct relic abundance of the wino-like neutalino as explicitly shown in our model. 
\clearpage
\begin{table}
\begin{center}
\begin{tabular}{|c|c||c|c|} \hline 
 &mass[GeV] & &mass[GeV]\\ \hline
$m_{\tilde{u}_1}$ &$1.3 \times 10^{6}$&$m_{\tilde{e}_1}$ &$1.2 \times 10^{6}$\\ \hline
$m_{\tilde{u}_2}$ &$1.2 \times 10^{6}$&$m_{\tilde{e}_2}$ &$9.3 \times 10^{5}$\\ \hline
$m_{\tilde{c}_1}$ &$9.1 \times 10^{5}$&$m_{\tilde{\mu}_1}$ &$1.0 \times 10^{6}$\\ \hline
$m_{\tilde{c}_2}$ &$8.9 \times 10^{5}$&$m_{\tilde{\mu}_2}$ &$9.0 \times 10^{5}$\\ \hline
$m_{\tilde{t}_1}$ &$7.0 \times 10^{5}$&$m_{\tilde{\tau}_1}$ &$9.0 \times 10^{5}$\\ \hline
$m_{\tilde{t}_2}$ &$4.1 \times 10^{5}$ &$m_{\tilde{\tau}_2}$ &$8.6 \times 10^{5}$\\ \hline
$m_{\tilde{d}_1}$ &$1.3 \times 10^{6}$ &$m_{\tilde{\nu}_{e_1}}$ &$1.2 \times 10^{6}$\\ \hline
$m_{\tilde{d}_2}$ &$1.3 \times 10^{6}$ &$m_{\tilde{\nu}_{e_2}}$ &$1.4 \times 10^{5}$\\ \hline
$m_{\tilde{s}_1}$ &$1.3 \times 10^{6}$ &$m_{\tilde{\nu}_{\mu_1}}$ &$1.0 \times 10^{6}$\\ \hline
$m_{\tilde{s}_2}$ &$8.9 \times 10^{5}$ &$m_{\tilde{\nu}_{\mu_2}}$ &$1.4 \times 10^{5}$\\ \hline
$m_{\tilde{b}_1}$ &$6.7 \times 10^{5}$ &$m_{\tilde{\nu}_{\tau_1}}$ &$8.6 \times 10^{5}$\\ \hline
$m_{\tilde{b}_2}$ &$4.1 \times 10^{5}$ &$m_{\tilde{\nu}_{\tau_2}}$ &$1.4 \times 10^{5}$\\ \hline
$m_{\tilde{\chi}_4^0}$ &$7.3 \times 10^{5}$&$m_{\tilde{\chi}_1^{\pm}}$ &377\\ \hline
$m_{\tilde{\chi}_3^0}$ &$7.3 \times 10^{5}$&$m_{\tilde{\chi}_2^{\pm}}$ &$7.3 \times 10^{5}$\\ \hline
$m_{\tilde{\chi}_2^0}$ &1227& $m_{3/2}$&$1.4 \times 10^5$ \\ \hline
$m_{\tilde{\chi}_1^0}$ &377& $m_{h}$&$125.5$ \\ \hline
$M_3$ & 3896 & &\\ \hline
 \end{tabular}
\end{center}
\caption{The mass eigenvalues of sparctile spectra, the Higgs boson mass $m_h$, the gravitino mass $m_{3/2}$, and the gluino mass $M_3$. 
The subscripts indicate the mass eigenvalues of sparticles, that is, the up ($\tilde{u}$), charm ($\tilde{c}$), 
top ($\tilde{t}$), down ($\tilde{d}$), strange ($\tilde{s}$), bottom ($\tilde{b}$) 
squarks, the scalar electron ($\tilde{e}$), muon ($\tilde{\mu}$), 
tauon ($\tilde{\tau}$), neutrino ($\tilde{\nu}$), the neutralino ($\tilde{\chi}$) 
and the chargino ($\tilde{\chi}^\pm$).}
\label{tab:spectra}
\end{table} 

\clearpage
\section{Conclusion}
\label{sec:con}
We have investigated the high-scale SUSY breaking scenario with the wino dark matter in the framework of $5$D SUGRA on $S^1/Z_2$ 
in which the successful moduli inflation can be realized~\cite{Abe:2014vca}. 
The systematic studies on the particle phenomenology as well as the cosmology are quite important to discuss the thermal history 
of the universe. Especially, the non-thermal dark matter abundance is highly model-dependent. 

In our model, the moduli stabilization as well as the moduli inflation have been achieved in a supersymmetric way, so that the gaugino 
masses are not generated at the tree-level. 
Even if we consider the SUSY breaking field, the leading contributions to the gaugino masses are anomaly mediation, 
whereas the gravity mediation determines the other soft SUSY-breaking terms which are controlled by the extra $U(1)$ symmetries broken by the orbifold projection. 
Then, the obtained sparticle spectra are similar to those of pure gravity mediation~\cite{Ibe:2011aa} 
and the dark matter candidate is the wino LSP whose abundance is determined by the thermal and non-thermal processes after the inflation. 
Since the inflaton and moduli are stabilized at their supersymmetric minimum, their decay into the gravitinos are suppressed by the almost vanishing F-terms of them. 
As a result, the non-thermal productions of the wino LSP produced by these fields decays are negligible. 

By contrast, the gravitinos are generated by the thermal bath and the decay of SUSY breaking field, and the wino LSP is produced by the gravitino decays hereafter. 
Although the relic abundance of the wino LSP is the sum of the thermal and the non-thermal abundances, 
it is found that the non-thermal process dominates the dark matter abundance consistent with the recent Planck data~\cite{Ade:2015lrj} even when the thermal abundance of it is negligible.

\subsection*{Acknowledgement}
The author would like to thank H.~Abe for useful discussions and comments. 
H.~O. was supported in part by a Grant-in-Aid for JSPS Fellows 
No. 26-7296.

\end{document}